\documentclass[journal]{IEEEtran}
\ifCLASSINFOpdf
\else
\fi
\usepackage{amsmath}
\usepackage{algorithmic}

\usepackage{array}
\usepackage{eurosym}

\newcommand{\RNum}[1]{\uppercase\expandafter{\romannumeral #1\relax}}
\usepackage{calc}
\usepackage{amssymb}
\usepackage{graphicx}
\usepackage{color}

\DeclareMathOperator*{\argmin}{arg\,min}

\newcommand{\scenario}[1]{Z}

\newcommand{\actionm}{u_m}

\newcommand{\actions}{u_s}
\newcommand{\actionrt}{u_\RT}
\newcommand{\actm}{u_m}

\newcommand{\da}{\text{da}}

\newcommand{\RT}{\text{RT}}

\newcommand{\s}[1]{S_{#1}}

\newcommand{\actionSetm}{\mathcal{U}_m}

\newcommand{\prob}[1]{\mathbb{P}\left\{#1\right\}}

\makeatletter
\newcounter{parentalgorithm}

\usepackage{subcaption}

\usepackage{textpos}
\usepackage{algorithm,algorithmic}
\usepackage{makecell}

\newcommand{\Rmnum}[1]{\expandafter\@slowromancap\romannumeral #1@}

\makeatother

\begin{document}
	%
	\title{Chance-Constrained Outage Scheduling using a Machine Learning Proxy}
	
	%
	%
	%

	
	\author{\IEEEauthorblockN{Gal Dalal\IEEEauthorrefmark{1},
			Elad Gilboa\IEEEauthorrefmark{1},
			Shie Mannor\IEEEauthorrefmark{1}, and
			Louis Wehenkel\IEEEauthorrefmark{2}} \\
		\IEEEauthorblockA{\IEEEauthorrefmark{1}Department of Electrical Engineering,
			Technion, Israel,
			\{gald,egilboa\}@tx.technion.ac.il, shie@ee.technion.ac.il}
		\IEEEauthorblockA{\IEEEauthorrefmark{2}Montefiore Institute -- Department of Electrical Engineering and Computer Science\\
			University of Li\`ege, L.Wehenkel@ulg.ac.be\\
		} 
	}
	
	%
	%

	\markboth{}%
	{Shell \MakeLowercase{\textit{et al.}}: Bare Demo of IEEEtran.cls for IEEE Journals}
	%



	\maketitle
	
	\begin{abstract}
		Outage scheduling aims at defining, over a horizon of several months {to years}, when different components needing maintenance should be taken out of operation. Its objective is to minimize operation-cost expectation while satisfying reliability-related constraints. We propose a distributed scenario-based chance-constrained optimization formulation for this problem. To tackle tractability issues arising in large networks, we use machine learning to build a proxy for {predicting outcomes of} power system operation processes in this context. On the IEEE-RTS79 and IEEE-RTS96 networks, our solution obtains cheaper and more reliable plans than other {candidates}.
	\end{abstract}
	


	%
	\IEEEpeerreviewmaketitle

\section{Introduction}	\label{chp:intro}

Outage scheduling is performed by transmission system operators (TSOs), as an integral part of asset management,  in order to carry out component maintenance and replacement  activities \cite{GARPUR5-1}. However, scheduling of the required outages for maintenance jobs is a complex task since it must take into account constrained resources ({e.g.} working crews, hours, and budget), increased vulnerability of the grid to contingencies during outages, and the impact of the scheduled outages on operations. {Moreover,} outage schedules, which are planned {on a mid-term scale} of several months {to years}, must also be robust with respect to uncertainties.

In this work, we present a general framework for assessing the impact of a given outage schedule on expected costs and system reliability, incurred while operating the system during {the schedule's} period. In addition, we formulate and solve a stochastic optimization program for optimally scheduling a list of required outages.  {To} do so, {we} take into account the smaller-horizon decision processes taking place during this time interval. These latter concern day-ahead operational planning and real-time operation. 

The complex dependence between the multiple time-horizons and the high uncertainty in the context of mid-term planning renders the corresponding assessment problem challenging. As demonstrated in \cite{dalal2016distributed}, solving an extensive amount of unit commitment (UC) problems to mimic short-term decision-making does not scale well to realistic grids, with thousands of buses or more. This is specifically burdensome {while simulating trajectories of months to years}. To deal with {this complexity} we propose to use machine learning to design a proxy that approximates short-term decision making, relieving the dependence of mid-term outcome assessment on {accurate} short-term simulations; ergo, allowing a tractable assessment method. Specifically, we replace exact UC computations with pre-computed UC solutions to problems with similar input conditions. See Section~\ref{sec:UCNN} for further details on the method.

When planning for future outages to enable maintenance, a certain reliability criterion is attempted to be satisfied at all future times. {Nowadays}, the common practice among TSOs is to consider the deterministic N-1 reliability criterion, while other probabilistic criteria are also being investigated \cite{karangelos2015probabilistic,GARPUR_2.2}. {To} make the system secured months in advance, the asset management operator should ideally assess whether each of the possible future scenarios {is} secure, by taking into account the coordination with day(s)-ahead and (intra)hourly operation. Since considering all possible realizations of future events is impractical, they must be approximated using sampled paths of future scenarios. {In this work, we thus also devise a sampling scheme that richly represents possible occurrences while being tractable.  We trust such methods are crucial for high-quality mid-term analysis.}

\subsection{{Related Work}}


Current practice in transmission system asset management offers  three main approaches: {time-based, condition-based, and reliability-centered preventive maintenance \cite{GARPUR5-2}. As for the academic literature, two popular trade-offs are i) increasing equipment reliability via maintenance while minimizing maintenance costs; and ii) minimizing the effect of outages on socio-economic welfare while satisfying operating constraints.}

In \cite{abiri2013two}, {the first above trade-off was considered in a two-stage formulation. The first stage schedules mid-term maintenance that imposes conditions on the second stage problem: short-term N-1 secured scheduling. By choosing a maintenance action per each time-interval, Weibull asset failure probability was analytically minimized. In \cite{jiang2004risk},  an analytic objective function was also designed. There, maintenance reduced cumulative risk of events such as overloads and voltage collapses, assuming known year-ahead generation and load profiles. The accumulated gain was negative during the actual maintenance, but positive afterwards due to its failure rate reduction.} 

{
As mentioned earlier, coordination with UC and economic dispatch may render the outage schedule assessment intractable, especially under security criteria. 
To overcome this, a coordination strategy between the different tasks was proposed in \cite{fu2007security}. There, mid-term planning over a deterministic 168-hour-long scenario minimized UC scheduling costs under changing network topology. In more recent work \cite{Marin2017schedulingproxy}, a greedy outage scheduling algorithm used Monte-Carlo simulations to assess the impact of outages on system operation. By mimicking experts heuristics for mid-term outage-scheduling, it enables long-term assessment of system development and maintenance policies.
}
\subsection{{Our Contribution}}
{Our contribution} is three-fold. First, we provide a new probabilistic mathematical framework that accounts for three entities involved in the multiple-horizon decision-making process; these are namely the mid-term, short-term and real-time decision makers. Our model captures their {hierarchy and formulates their coordination using an information sharing scheme, that limits each via partial observability.}

\begin{figure}
	\includegraphics[scale=0.49]{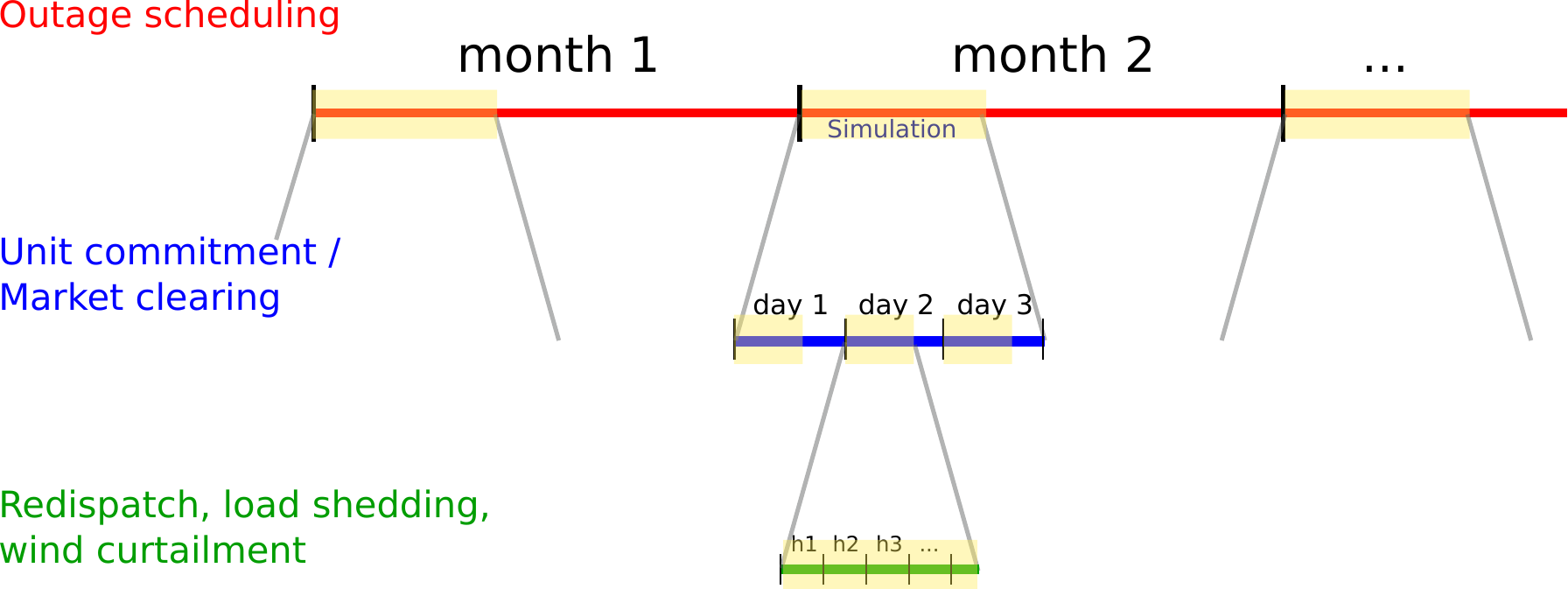}
	\caption{Our \emph{Bayesian hierarchical window sampling} is a scenario generation approach, {which interpolates between} sequential trajectory simulation and snapshot sampling. In each level of the hierarchy, a snapshot of future grid and environment conditions is sampled, and sequential simulation is performed from that point on, for a limited time window. }  \label{fig:simulation}
\end{figure}

Second, we introduce the novel concept of integrating a component that instantly predicts approximated short-term decision outcomes; we refer to it as a \emph{proxy}. We do so with a well-known machine learning algorithm, \emph{nearest neighbor}. We thus enable a critical reduction in computation time, which turns the table and deems ֿ{large-scale multiple time-horizon simulation-based assessment tractable.}

Third, we devise a scenario-based optimization methodology. {It builds on our} scenario generation approach, \emph{Bayesian hierarchical window sampling}, which {interpolates between} sequential trajectory simulation and snapshot sampling while accounting for coordination between the three decision layers; see Fig.~\ref{fig:simulation}. Using it, we solve {our stochastic chance-constrained outage scheduling formulation for IEEE-RTS79 and IEEE-RTS96 with distributed computing and show promising results.}

\section{Mathematical Problem Formulation} \label{sec:problem_formulation}

In our problem setting, the TSO lists necessary outages $\{o_{1}, \ldots , o_{n_{o}}\}$, each one defined by a duration, a budget of crew requirements, and a specific component to be taken out of operation for maintenance. A specific component can be required to undergo more than a single outage. The outage scheduling horizon (e.g. several months, or a couple of years) is split into $T$ hourly time-steps. The decision variable $\actionm$ is a vector of length $n_{o}$, with $u_m (i) \in {\cal T}_{m}$ denoting the starting time of outage $o_{i}$ and where ${\cal T}_{m}$ is a given set of candidate outage moments (e.g. monthly steps); we denote by ${\cal U}_{m} = ({\cal T}_{m})^{n_{o}}$ the set of candidate outage plans. We formulate the mid-term stochastic optimization program in \eqref{eq:fullMidTermFormulation}, where the goal is to minimize expected future operating costs by adopting an optimal planned outage schedule $\actionm$.
\begin{subequations}\label{eq:fullMidTermFormulation}
	\begin{align}
	& \min_{\actionm \in {\cal U}_{m}} \sum_{t=1}^{T} \mathbb{E}_{{\s{t} \in \scenario{m}}} \left\{{C}(\s{t},\actionm, u_s^*, u_\RT^*)\right\} \label{eq:midTermObj}  \\
	& ~\text{s.t.} ~~ \mathbb{P}\left\{r(\scenario{m}, \actionm, u_s^*, u_\RT^*) \ge \underbar{$r$} \right\} \geq 1- \alpha_\text{r} \label{eq:reliabilityCC}\\
	& ~\quad ~~ \mathbb{P}\left\{\text{LS}(\scenario{m}, \actionm, u_s^*, u_\RT^*) \leq \overline{\text{LS}} \right\} \geq 1- \alpha_\text{LS} \label{eq:LSCC}\\
	&\quad\quad h(\actionm) \le 0  \label{eq:midTermFeasibility}\\
	&\quad\quad  u_s^* = \argmin_{u_s \in \mathcal{U}_s(u_m)}   C_s(u_s,y_s,u_m) \label{eq:optimalS}\\
	& \quad \quad u_\RT^* = \argmin_{u_\RT \in \mathcal{U}_\RT(u_s^*, \actionm)}  C_\RT(u_\RT,y_\RT,u_s^*, \actionm).  	\label{eq:optimalRT}
	\end{align} 
\end{subequations}
{This formulation's components are explained as follows.}

\subsection{Objective}
The objective in \eqref{eq:midTermObj} is the aggregated expected cost of operational decisions summed over the evaluation horizon. The expectation is w.r.t. the distribution of the uncertain future conditions of the grid encased in and denoted by stochastic {exogenous} scenario $\scenario{m}$. Scenario $\scenario{m}=(\s{1},\dots,\s{T})$ is a series of states $\s{t}$, {which are introduced in more detail} in Section~\ref{sec:state-space}. When making decisions in the mid-term time-horizon, one must take into account the smaller-horizon decisions that take place during it. In this work, the smaller-horizon decisions considered are short-term (day-ahead) operational planning $\actions \in \mathcal{U}_s(u_m)$ and real-time control $u_\RT \in \mathcal{U}_\RT(u_s)$. {Each of the sets of candidate smaller-horizon decisions $\mathcal{U}_s(u_m),~\mathcal{U}_\RT(u_s, \actionm)$ is function of the decisions that are taken higher in the hierarchy.} 

In our work, a real-time decision $u_{\RT}^*$ defines a vector of redispatch values for each redispatchable generator, wind curtailment values for each wind generator, and load shedding values for each bus, and is determined by minimizing the cost $C_{\RT}$ of deviation from the day-ahead market schedule $u_s^*$. The latter is determined by minimizing a day-ahead {objective} $C_{s}$. {Lastly, we choose the function {$C(\s{t},\actionm, u_s^*, u_{\RT}^*)$} to only account for the real-time operating costs. Specifically, it is identical to $C_\RT,$ apart for the load shedding cost. This is because load shedding is already addressed via \eqref{eq:LSCC}.  However, the formulation is general, and may in principle also incorporate real-time load shedding, as well as market surplus and day-ahead reserve purchase costs, if deemed important.}

\subsection{Reliability and Load Shedding Chance-Constraints}
{
To maintain the generality and flexibility of our model while respecting its probabilistic framework, we define a reliability metric that is independent of smaller time-horizon {specificities. It allows for} equitable comparison between different maintenance strategies and operation policies.
Inspired by the common N-1 criterion used in the industry, we adapt it to our probabilistic setup. Namely, we consider the system's ability to withstand any contingency of a single component. We thus define reliability as the portion of contingencies under which the system retains safe operation, which, practically, we measure via AC power flow convergence.
}

{
For this, denote by ${\mathcal N_{-1}}(S_t, \actionm )$ the N-1 contingency list and by  $r(\s{t}, \actionm, u_s^*, u_\RT^*) \in [0,1]$ the real-time reliability, which for brevity we also denote by $r(\s{t}, \actionm )$. The latter is calculated for given state $\s{t}$ and is dependent of current topology, dictated by $\actionm$:
\begin{equation*}
r(\s{t}, \actionm ) = \frac{1}{|{\mathcal N_{-1}}(S_t, \actionm )|}\sum_{c \in {\mathcal N_{-1}}(S_t, \actionm )}\mathbb{I}_{[\text{PF}(c,S_t, \actionm)]},  
\end{equation*}
where $\mathbb{I}_{[\text{PF}(c,S_t, \actionm)]}$ equals $1$ if a feasible ACPF solution exists during contingency $c \in {\mathcal N_{-1}}$, and $0$ otherwise. 
Also, let 
\begin{equation}
\label{eq:reliability}
r(\scenario{m} , \actionm, u_s^*, u_\RT^*) = \frac{1}{T} \sum \limits_{\s{t} \in \scenario{m}}  r(\s{t}, \actionm);
\end{equation} 
i.e., the average success rate for scenario  $\scenario{m}$.
}

{
In similar fashion, let $\text{LS}(\s{t}, \actionm )$ be the total load shed in state $\s{t}$, as determined by the real-time decision $u_{\RT}^*$, and  $\text{LS}(\scenario{m}, \actionm , u_s^*, u_\RT^*) = \frac{1}{T} \sum \limits_{\s{t} \in \scenario{m}}\text{LS}(\s{t}, \actionm );$ i.e., the average amount of load shed during scenario $\scenario{}$.
}
Based on the {these definitions}, the chance-constraints in \eqref{eq:reliabilityCC}-\eqref{eq:LSCC} ensure that the average  reliability remains above a minimal value $\underline{r}$ and that the average load shedding remains below a maximal value $\overline{\text{LS}}$, with respective probabilities $1- \alpha_\text{r}$ and $1- \alpha_\text{LS}$. Based on reasonable achievable values for our specific test-case modifications listed in Section~\ref{sec:experiment_results}, throughout this work we set ${\underline{r} = 0.8,\overline{\text{LS}}=0.5\% \mbox{ of overall load capacity}} ,\alpha_\text{r}=0.05,~\alpha_\text{LS}=0.05.$ 

The reason for explicitly incorporating the two chance constraints together is to ensure both high reliability and low load shedding at the same time, as these two obviously  trade-off. We further relate to this trade-off in {Section~\ref{sec:experiment_results}}.
\subsection{Feasibility Constraints}
Maintenance feasibility constraints $h(\actm)$ in \eqref{eq:midTermFeasibility} define which maintenance schedules are feasible, e.g., cannot maintain more than two assets per month.

\subsection{Coordination with Smaller-Horizon Subproblems}
Lastly, the constraints in \eqref{eq:optimalS}-\eqref{eq:optimalRT} ensure coordination between mid-term and smaller-horizon decisions. The informational states $y_s$ and $y_\RT$ appearing as arguments {of $C,~C_s,~C_\RT$} depict the partial information revealed to the respective decision makers in these time-horizons; further details on the notion of informational states are given in Section ~\ref{sec:informational_state}.

\section{Probabilistic Model Components}	
\label{sec:components}
In this section, we {elaborate on our probabilistic decision making model}. We define a state-space representation encapsulating {all exogenous uncertain information} and the decision makers' limited access to this information. Our model is generic and can be adapted for additional uncertain factors.
\subsection{State-Space}	
\label{sec:state-space}
{
\textit{State} $\s{t}\in \mathcal{S}$ captures all exogenous uncertain information at time $t$, required to make informed decisions in all considered time-horizons.}
Let $n^l_t,~n^b,~n^g_d,~n^g_w$ respectively be the number of transmission lines, buses, dispatchable generators, and wind generators in the network. 
The state $\s{t}$ is defined as the following tuple:
\[\s{t} = (J_t,\hat{W}_{\da},\hat{D}_{\da},W_t,D_t,\text{top}_t), \mbox{where}\]
\begin{itemize}
	\item $J_t \in \mathbb{R}^2$ is the seasonal weather factor, determining the intensity of demand and wind generation. This variable changes monthly, with values drawn around a mean profile corresponding to typical  seasonal trends.
	\item $\hat{W}_{\da} \in \mathbb{R}_+^{n^g_w \times T_{\da}}$ is the day-ahead wind generation forecast, where
	$T_{\da}$ is the day-ahead planning horizon {(24 in our simulations)}. Notice that variables with subscript '$\da$' {remain} fixed for time periods of length $T_{\da}$, and are updated each $T_{\da}$ time-steps.
	\item $\hat{D}_{\da} \in \mathbb{R}_+^{n^b \times T_{\da}}$ is the day-ahead load forecast.
	\item $W_t \in \mathbb{R}_+^{n^g_w}$ is the realized wind generation at time-step $t$. It is {assumed fixed during the intra-day interval (1 hour).}
	\item $D_t \in \mathbb{R}_+^{n^b}$ is the realized load at time-step $t$.
	\item $\text{top}_t \in \{0,1\}^{n^l}$ is the network {transmission line} topology at time-step $t$,  { as determined by exogenous events. Entry $\text{top}_t(i_l) = 0$ indicates line $i_l$ is offline at time $t,$ due to a random forced outage.}
\end{itemize}

\subsection{Informational State{s}}
\label{sec:informational_state}

{
Decision makers with different time-horizons are exposed to different degrees of information; i.e.,  the higher the decision's temporal resolution, the more state variables are realized at the time of the decision. We formulate these granularities via \emph{informational states} as follows. Denote $\s{t}^{1:k}$ to be $\s{t}$'s sub-vector containing entries $1$ to $k$. Let 
$y_s=\s{t}^{1:3}=(J_t,\hat{W}_{\da},\hat{D}_{\da})$ and $y_\RT=\s{t}$; these are respectively the short-term and real-time informational states. When performing his decision, the short-term planner is exposed to  $y_s,$ which  carries the realizations of the day-ahead generation and load forecasts. Notice he is also exposed to the higher-level mid-term decision $\actionm$; however, we do not model it as a part of the informational state as it is not exogenous. As for the real-time operator, he is exposed to realized values of all state entries, i.e. $y_{\RT}$, and is similarly informed of the higher level decisions  $\actionm$ and $\actions$.} 

{For completeness, we also define the mid-term informational state $y_m=\s{t}^1=J_t,$ even though it does not appear directly in \eqref{eq:fullMidTermFormulation} (it does appear later for scenario generation purposes).} Notice that in our work $\actionm$ is an open-loop mid-term strategy,  so that we do not use $y_m$ to revise mid-term decisions.

\subsection{Smaller-horizon Formulations} \label{sec:action_space}
Our formulation contains three hierarchical levels of decision making, namely mid-term outage scheduling, short-term {day(s)-ahead planning, and (intra)hourly} real-time control. We often refer to the short-term and real-time problems as the \emph{inner subproblems}. We now present the candidate decisions in these latter. 
\subsubsection{Short-term Formulation}
The formulation
\begin{equation} \label{eq:optimals}
u_s^* = \argmin_{u_s \in \mathcal{U}_s(u_m)} \quad  C_s(u_s,y_s,u_m),
\end{equation} {which also appears in \eqref{eq:optimalS}, is set in this work to be UC. As explained in Section~\ref{sec:problem_formulation}, we choose the cost $C$ in \eqref{eq:midTermObj} to be solely based on real-time realizations and decisions. Thus, the UC cost here is not to be minimized by the mid-term planner{; rather,} the UC solution is plugged into the real-time problem for setting commitment constraints and redispatch costs reference. 
Notice the UC formulation depends on day-ahead forecasts of wind power and load $\hat{W}_{\da},\hat{D}_{\da}$. These are parts of the informational state $y_s$, to which the decision maker is exposed when facing his day-ahead planning {decision}. The feasible action-space  $\mathcal{U}_s(u_m)$ in \eqref{eq:optimals} depends on the {topology set by the} mid-term decision $\actionm$, and it may also embody a reliability criterion of choice, e.g. N-0 or N-1. 
\subsubsection{Real-time Formulation}
{
The formulation 
\begin{equation} \label{eq:optimalrt}
u_\RT^* = \argmin_{u_\RT \in \mathcal{U}_\RT(u_s^*,{\actionm})} \quad  C_\RT(u_\RT,y_\RT,u_s^*, \actionm),
\end{equation} which also appears in \eqref{eq:optimalRT},}
follows the UC solution in \eqref{eq:optimals} as a baseline. 
It is solved sequentially for each hour, where each solution at time $t$ is fed to the next one at time $t+1$ to incorporate temporal constraints. {Similarly as in \eqref{eq:optimals}, a reliability criterion of choice may be ensured via the definition of the set $\mathcal{U}_\RT(u_s^*, \actionm).$}


\subsection{Scenario Generation} \label{sec:scenario_space}
{To solve \eqref{eq:fullMidTermFormulation} in the face of exogenous uncertainties and the intricate} interaction between these uncertainties, we rely in this work on scenario-based simulation \cite{dembo1991scenario}. 
{
\label{sec:scenario evaluation}
Existing literature on scenario generation splits into two main categories.
 The first is full-trajectory simulation \cite{jiang2004risk}, where (intra)hourly developments are simulated as a single long sequence. In our mid-term problem that spans over a whole year, such an approach will result in high variance and possibly necessitate an intractable number of samples to produce a decent evaluation of scenario costs.
 The second category of approaches is based on snapshot sampling of static future moments \cite{pineda2016impact}. The main issue with this methodology is the loss of temporal information.}

In light of this, we introduce a new scenario generation approach, \emph{Bayesian hierarchical window scenario sampling}, which is a hybrid of the two aforementioned methodologies, aimed at mitigating the disadvantages of each of them. 
 Visualized in Fig. \ref{fig:scenarioApproximation}, our sampling process begins with drawing monthly parameters for wind and load intensity, i.e., drawing sequence $\{y_m^{(t_m)}\}$ from transition probability $\prob{y_m^{(t_{m+1})}|y_m^{(t_m)}},$ {where $t_m$ is a monthly time index}. Then, we draw $N_s$ replicas of $W_s$\footnote{The choice of the window lengths $W_s$ and $W_\RT$ controls the level of interpolation between the completely sequential scenario sampling of all $T$ time steps, and the alternative completely static approach of solely sampling snapshots of states, with no temporal relation between them. Essentially, they arbitrate between the bias and variance of the sampling process. Full trajectory sampling has low bias but high variance, while static snapshot sampling lowers the variance, though it introduces bias due to its simplicity and choice of times of snapshots.} consecutive days; this results in sequences $\{y_s^{(t_s)}\}$ drawn from transition probability $\prob{y_s^{(t_{s+1})}|y_s^{(t_s)},y_m^{(t_m)}},$  {where $t_s$ is a daily time index}. Lastly, per each such day, we draw $N_{\RT}$ replicas of $W_{\RT}$ consecutive hours and form sequences $\{y_{\RT}^{(t)}\}$ from transition probability $\prob{y_{\RT}^{(t+1)}|y_{\RT}^{(t)},y_s^{(t_s)},y_m^{(t_m)}}$. Having realizations of day-ahead forecasts in $y_s$ and their corresponding hourly realizations in $y_{\RT}$, we can respectively solve the {daily  and hourly inner subproblems}. Based on the incurred costs, we are able to evaluate the {scenarios' accumulated costs} per each month in a parallel fashion.

\begin{figure}
	\centering
	\vspace*{-7mm}\includegraphics[scale=0.38]{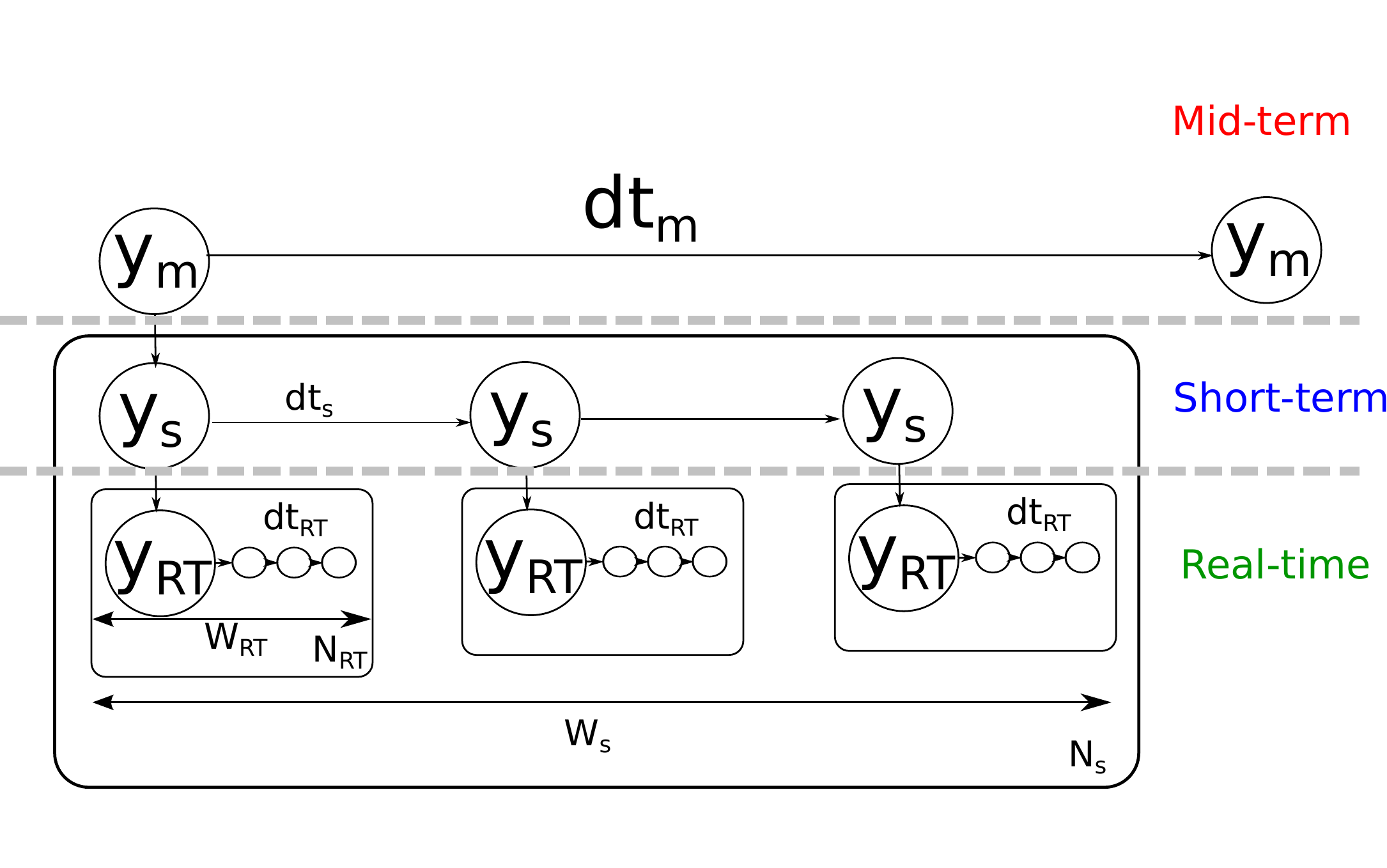}
	\caption{{Our Bayesian hierarchical window scenario sampling approach for scenario generation relies on a conditional factorization of the state to its three informational state.}}
	\label{fig:scenarioApproximation}
\end{figure}

In supplementary material \cite{dalal2017chancesup} we provide more details on {the probabilistic models used in} this sampling approach.

\begin{figure*}
	\centering
	\includegraphics[scale=0.3]{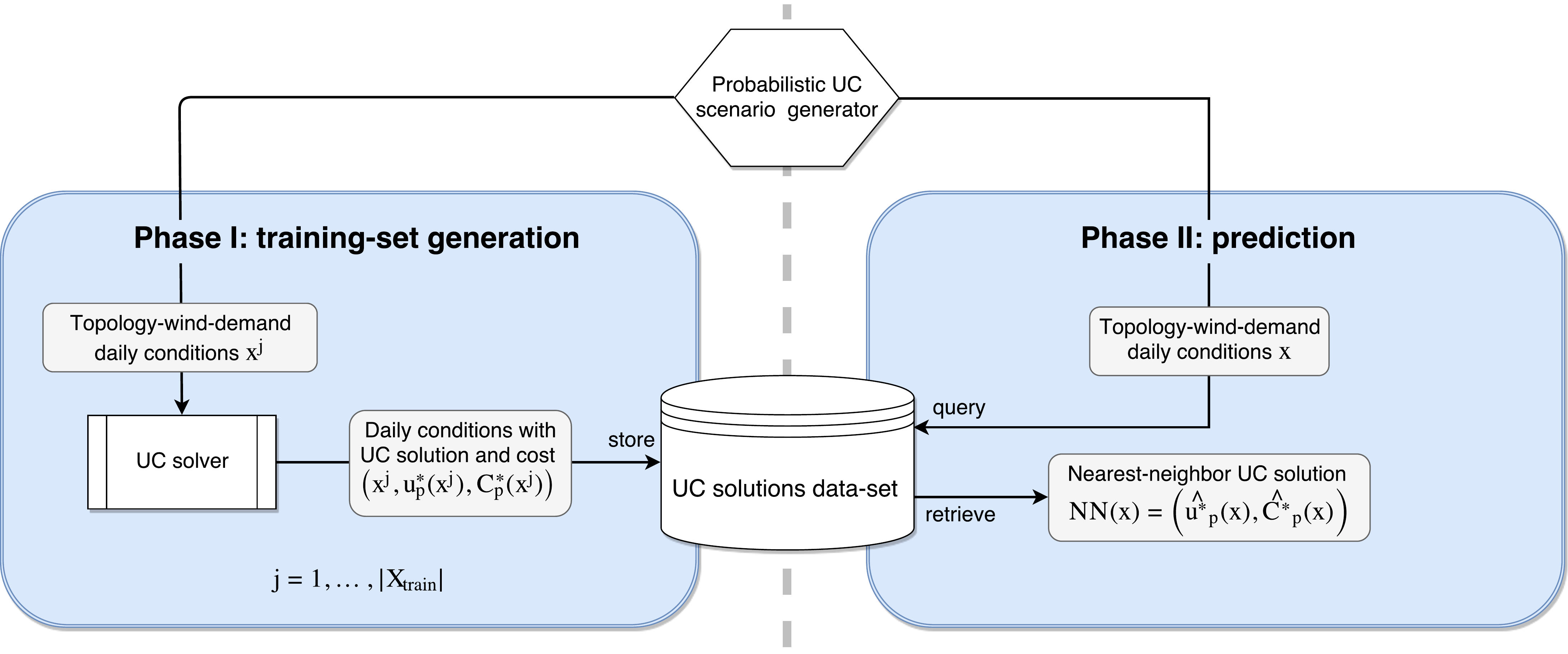}
	\caption{UCNN algorithm diagram. In an initial phase, multiple UC {inputs} are generated, solved, and then stored along with {their} solutions and costs to create a diverse dataset, also referred to as \emph{training set}. In a second phase, when a new UC problem instance is received, an approximate UC solution is obtained by finding a nearest-neighbor among the existing solutions in the dataset; i.e., a pre-solved similar problem instance. This is to replace the usage of the computationally expensive UC solver.}
	\label{fig:block diagram}
\end{figure*}

\section{Machine Learning for a Short-Term Proxy\label{sec:UCNN}}

As mentioned earlier, in this work we utilize machine learning {to build a short-term proxy. Thus, we replace exact solutions  of the multiple UC problem instances, originating in \eqref{eq:optimalS},} with their quickly-predicted approximations.  We use a well-known machine learning algorithm: \emph{nearest neighbor classification} \cite{cover1967nearest}{; we} thus call it UCNN. 
The methodology relies on a simple concept: creating a large and diverse data-set that contains samples of the environment and grid conditions along with their respective UC solutions. {Then}, during the assessment of an outage schedule, instead of solving the multiple UC problem instances required to simulate decisions taken, we simply choose among the already pre-computed UC solutions. The solution used is the one with the closest conditions to the environment and grid conditions of the current UC problem needed to be solved; hence the phrase nearest neighbor.
We note that to confidently obtain high-quality approximate solutions, we generate the data-set so as to cover all relevant topologies {that might be encountered during prediction}. In our context, this implies a data-set that is $O(2^{|{\cal L}|})$, where ${\cal L}$ is the set of {components} for which outages ought to be performed. In general, this combinatorial dependence is not necessarily compulsory. Nevertheless, the question of accuracy degradation with smaller data-sets and more efficient data-set compositions are subject to future work.

The method's strength stems from the fact that  during the optimization process {\eqref{eq:fullMidTermFormulation}}, which is based on multiple {outage} schedule assessments, UC problem instances are often  similar to previously computed ones. Therefore, instead of repeating the expensive process of obtaining these solutions during the assessment of a single scenario and across different scenarios, we utilize samples created ex-ante as representatives of similar conditions. The initial creation of the dataset is a slow process which can either be done offline or online, i.e., by continually adding new solutions to the dataset as they become available. For the experiment described in this section, a dataset of $5000$ UC problem instances was created.
After obtaining this initial dataset, UCNN reduces computation time in several orders of magnitude, with relatively little compromise in quality \cite{dalal2016unit}.  A diagram visualizing the method is given in Fig.~\ref{fig:block diagram}.

In addition to the {direct UC approximation comparison} in \cite{dalal2016unit}, we examine the {resulting accuracy of outage scheduling assessment when using UCNN to solve the short-term subproblem instead of exact UC computations}. To do so, we generate four arbitrary outage schedules under the configuration given in Section~\ref{sec:experiment_results}. Then, for each of these schedules, we present in Fig. \ref{fig:UCNN_schedules} means and standard deviations of several metrics in our simulation across the year's progress. These metrics are i) day-ahead operational costs and ii) load shedding amounts, taken from the short-term UC simulation. Additionally, they include the real-time values of iii) reliability as defined in \eqref{eq:reliability} and vi) real-time operational costs. In all of these plots, the red curve is of an empty, no-outage schedule given as a baseline, evaluated using exact UC simulation; the blue and green curves are respectively based on exact UC and UCNN simulations of the arbitrary outage schedules. The persistent proximity of the green curve to the blue demonstrates the low approximation error when using UCNN, as it propagates to the four inspected metrics during the simulation.

\begin{figure}
	\centering
	\includegraphics[scale=0.26]{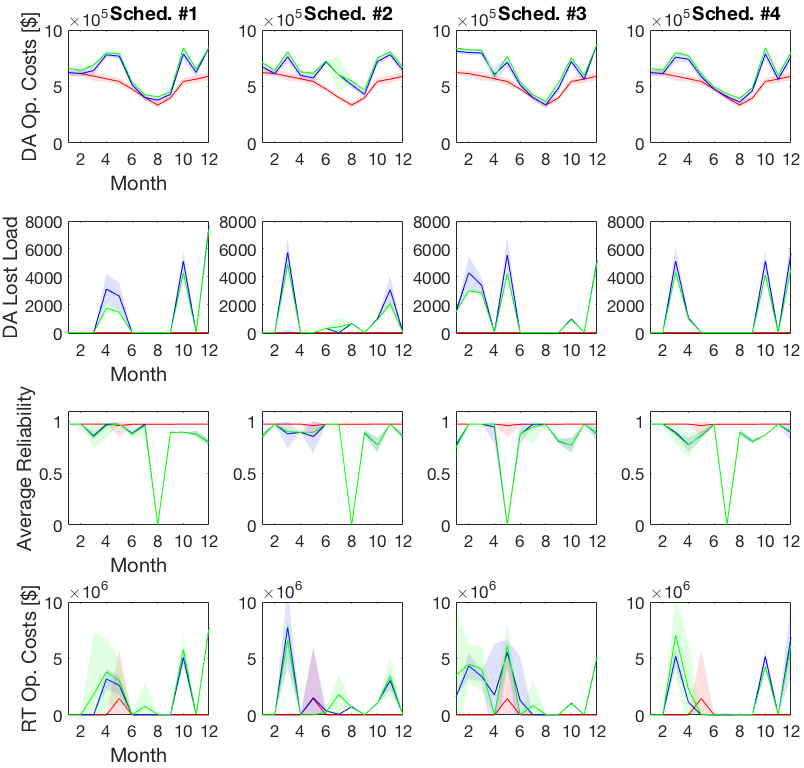}
	\caption{Low proxy approximation errors demonstrated for IEEE-RTS79. Plotted are {monthly} costs of no-outage schedule as a reference (red) and four arbitrary outage schedules, evaluated with exact UC (blue) and UCNN  (green).}\vspace*{-3mm}
	\label{fig:UCNN_schedules}
\end{figure}

\section{Distributed Simulation-Based Cross Entropy} \label{sec:cross_entropy}
Problem \eqref{eq:fullMidTermFormulation} is a non-convex combinatorial stochastic optimization program, with inner mixed integer-linear programs (MILP). 
{As such}, it is too complex for the objective and constraints to be expressed in explicit analytic form and for the program to be solved using gradient-based optimization{. Deriving} the gradient estimator is a challenging problem and requires large quantities of generated data \cite{nesterov2017random}. Furthermore, gradient-based approaches would preclude the option of utilizing smaller-horizon machine learning proxies such as our UCNN. For this reason, we choose a gradient-free simulation-based optimization approach. It is performed with distributed Monte-Carlo sampling, where multiple solutions in $\actionSetm$ are being assessed in parallel on multiple servers. Each month of such solution assessment is itself simulated in parallel. 

Our algorithm of choice is Cross Entropy (CE) \cite{de2005tutorial}. It is a randomized percentile optimization method that has shown to be useful for solving power system combinatorial problems \cite{ernst2007cross}. In CE, some parametric distribution is maintained over the solution space. It iteratively performs consecutive steps of i) drawing multiple candidate solutions (outage schedules) and evaluating their objective value, followed by ii) updating the parametric solution distribution based on the top percentile of samples. {More details on our CE implementation are given in the supplementary material \cite{dalal2017chancesup}.} Constraint satisfaction is ensured in the following way: Chance-Constraints~\eqref{eq:reliabilityCC}-\eqref{eq:LSCC} are evaluated empirically and their violation is penalized with increasing-slope barrier functions; Feasibility Constraint~\eqref{eq:midTermFeasibility} is ensured via the structure of the CE parametric distribution; Inner Constraints~\eqref{eq:optimalS}-\eqref{eq:optimalRT} are ensured {via the solvers used for simulating them.}

\section{Experiments} \label{sec:experiment_results}
We run our experiments on a Sun cluster with Intel Xeon(R) CPUs  $@2.53$GHz, containing {a total of} 300 cores, each with 2GB of memory. All code is written in Matlab \cite{matlab}. {In each iteration of the CE algorithm, we assess the objective and constraint values of $75$ drawn outage schedules in parallel, while also parallelizing the simulation of each of the $12$ months. The simulation parameters introduced in Section~\ref{sec:scenario_space}, depicting daily and hourly {trajectory} length and multiplicity, are $W_s=3~,N_s=4,~W_{\RT}=24,~N_{\RT}=2$.} This totals a year-long trajectory which is sampled $3$ times.

\subsection{Model Assumptions}
In our experiments we set the horizon in \eqref{eq:fullMidTermFormulation} to be one year, i.e., $T=24\times 365.$ {We consider transmission line outages with} monthly candidate outage moments, i.e. ${\cal T}_{m} = \{1,\dots,12\},$ where the duration of each outage is one month. 

As for the inner subproblems, the UC formulation in \eqref{eq:optimalS} is implemented as a multi-stage DCOPF constituting a MILP, where the {objective $C_{s}$ and decision variable $\actions$ compose of generation, start-up, wind-curtailment, and load shedding.}  The real-time formulation in \eqref{eq:optimalRT} is implemented similarly as an hourly DCOPF, where the {objective $C_{\RT}$ and decision variable $\actionrt$ compose} of redispatch, wind-curtailment, load shedding and re-commitment w.r.t. to the solution of \eqref{eq:optimalS}.  For both short-term and real-time, the reliability criterion used is N-0; i.e., no contingency list is {considered at the subproblem level}. Instead, {our probabilistic notion of N-1 resiliency is ensured via \eqref{eq:reliabilityCC}}. The rigorous formulations of subproblems \eqref{eq:optimalS}-\eqref{eq:optimalRT} are given in the supplementary material \cite{dalal2017chancesup}. {We model these optimizations with YALMIP \cite{YALMIP} and solve with CPLEX \cite{CPLEX}}. 

\subsection{Test-Cases and Outages}

In our simulation, we consider the IEEE-RTS79 and IEEE-RTS96 test-cases \cite{subcommittee1979ieee}. We adopt updated generator parameters from Kirschen et al. \cite{pandzic2013comparison}, namely their capacities, min-output, ramp up/down limits, min up/down times, price curves and start-up costs. {Peak loads and hourly demand means are based on data from the US published in \cite{UW_website}. Capacities and means of hourly wind generation are also based on real data, taken from \cite{UW_website}.   Based on these values, b}oth demand and wind processes are sampled from a multivariate normal distribution with standard deviation that is a fixed fraction of the mean  -- $0.02$ for demand and  $0.15$ for wind. Moreover, a monthly trend in demand and wind is governing the mean profiles \cite{ouammi2010monthly}.
Value of lost load is set to $VOLL=1000[\frac{\$}{MWh}]$, taken from \cite{dvorkin2015hybrid} and wind-curtailment price is set to $C_{\text{WC}}=100[\frac{\$}{MWh}]$, taken from \cite{loisel2010valuation}.

Additionally, we slightly modify the test-cases so as to create several 'bottleneck' areas to provide conditions for {variant outage schedule qualities}. In RTS79, these modifications include i) removal of transmission line between bus 1 and 2, and ii) shift of loads from buses 1 and 2 to buses 3 and 4, respectively. In the case of RTS96, the same exact modifications are replicated to all three zones.

Next, we specify the outage lists. For RTS79, it is composed of $13$ outages: $2$ outages per each of lines $\{2,3,4,5,25,26\}$ and $1$ outage for line $11.$ For RTS96, the list is composed of $30$ outages: $9$ per each zone plus $3$ for the interconnections. {Specifically}, in the first zone of RTS96 we have $2$ outages per each of lines $\{2,3,4,5\}$ and $1$ outage for line $11;$ these are replicated similarly to the equivalent lines in the second and third zones. The additional interconnection outages are $1$ per each of lines $\{12,119,120\}.$ The test-case modifications and outages are visualized in Fig. \ref{fig:case96_modifications}.



As explained in Section~\ref{sec:UCNN}, the UCNN proxy data-set size is linearly dependent on the outage combinations set. This implies a huge data-set that is $O(2^{3\times5+3})$ given the above-listed outages. 
To tackle this, in this work, when scheduling outages for RTS96, we assume that the year is partitioned into three periods of 4 months, and each of the three ``zone operators'' is exclusively allocated with distinct $4$  months to conduct his $9$ outages. This is enforced via the feasibility constraint \eqref{eq:midTermFeasibility}. As for the outages of the additional $3$ interconnections, those are {independent and free to be chosen to} any of the year's $12$ months. We thereby do away with the exponential dependence of UCNN's dataset complexity in the number of zones, i.e., reduce the $O(2^{3\times5+3})$ training set size to $O(3\times2^{5+3})$.

\begin{figure} 
	\includegraphics[scale=0.383]{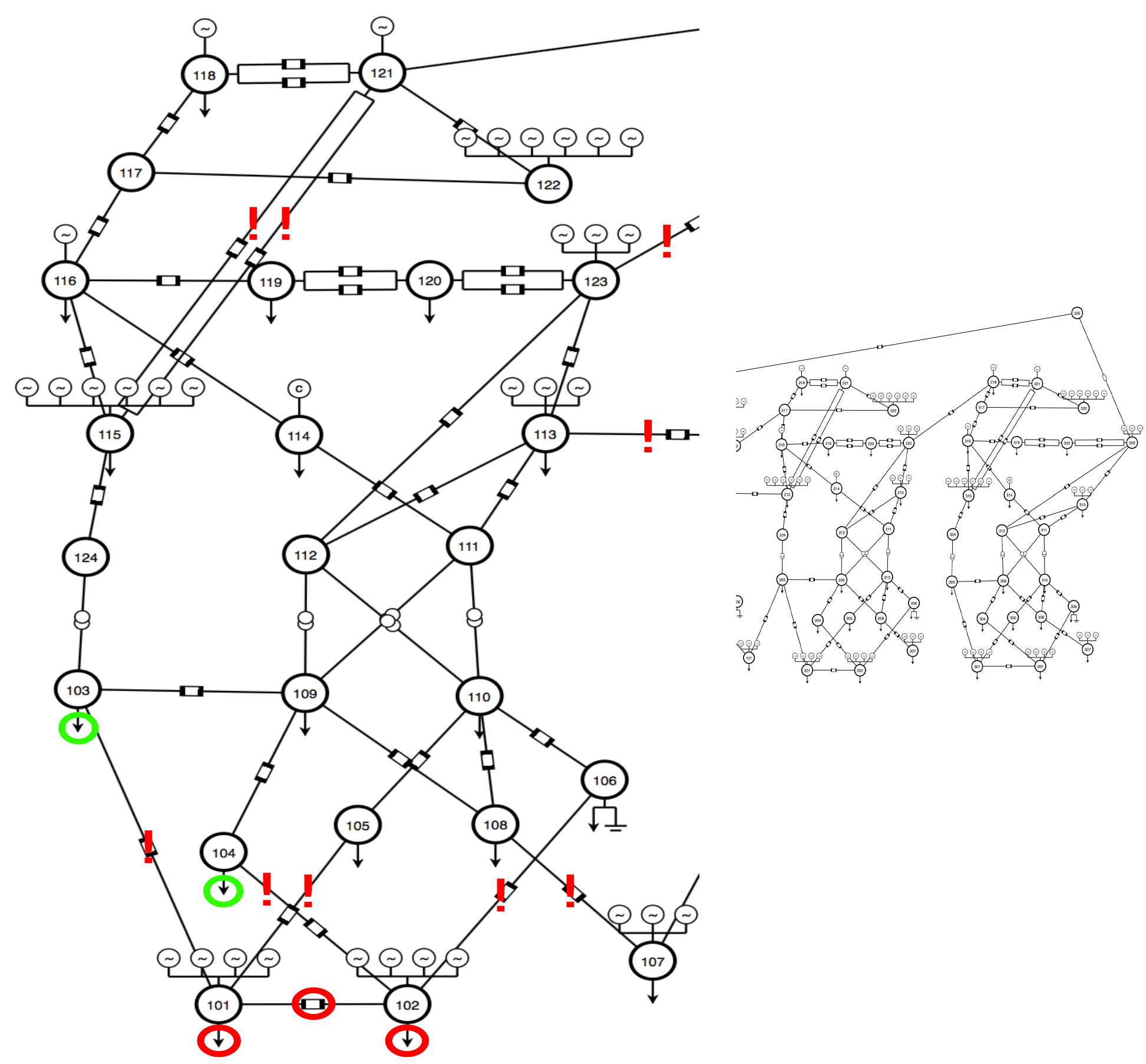}
	\caption{Modifications and target outages in the IEEE-RTS96 test-case (conducted per each of the zones, though plotted on one for simplicity). Red circles denote removal, green circles denote increase, and red exclamation marks denote candidate planned outage. {The top two outages on the left are not considered in RTS96, but only RTS79.} Furthermore, in RTS79 the same exact modifications are conducted for the single zone, and the outages do not include the interconnections.}
	\label{fig:case96_modifications}
\end{figure}

\subsection{Results}

Fig.~\ref{fig:convergence} exhibits the fast convergence as expected from CE {when solving \eqref{eq:fullMidTermFormulation}} for the two test-cases, along with intriguing differences between them. It gives the median with upper and lower quartiles of {the top CE percentile for} three metrics: operational costs from \eqref{eq:midTermObj} (redispatch, wind curtailment and unit re-commitment), average reliability from \eqref{eq:reliabilityCC}, and average load shedding from \eqref{eq:LSCC}. In both test-cases, operational costs significantly drop. As for the reliability and load shedding, a somewhat complementary behavior is observed for the two test-cases. The reliability in RTS79 starts off with high enough values, $83\%$, to satisfy its constraint \eqref{eq:reliabilityCC}, while the load shedding amount starts high and quickly drops to a satisfying level of $0.4\%$. The exact opposite happens for RTS96: reliability starts low and increases drastically to $83\%$, while load shedding values consistently remain low throughout the optimization process, stabilizing at $0.05\%$.

\begin{figure}
	\centering
	\begin{subfigure}[]{0.50\textwidth}
		\includegraphics[width=8.5cm,height=3.7cm]{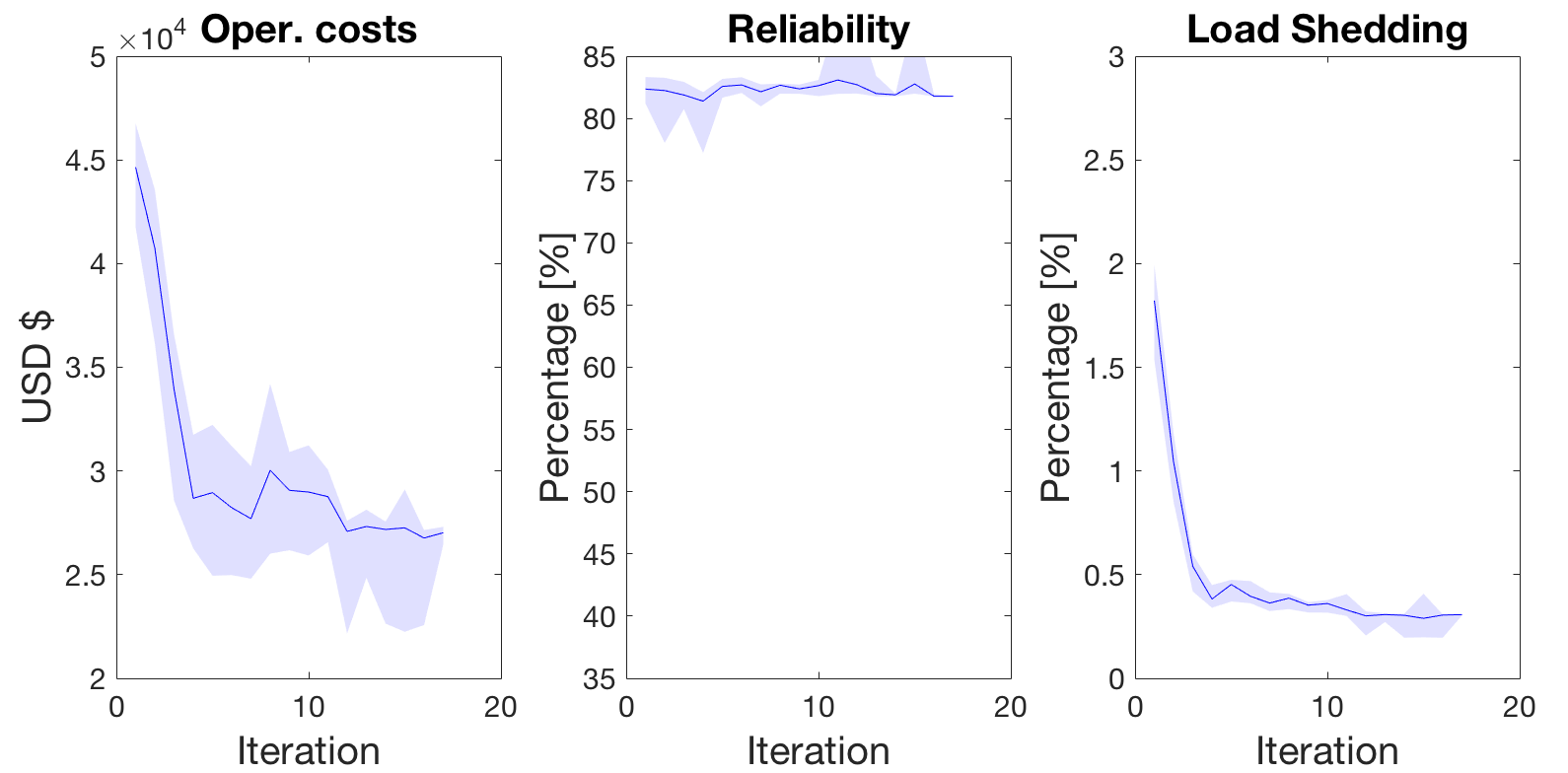}
		\caption{IEEE-RTS79}\vspace*{2mm}
	\end{subfigure}
	
	\begin{subfigure}[b]{0.5\textwidth}
		\includegraphics[width=8.5cm,height=3.7cm]{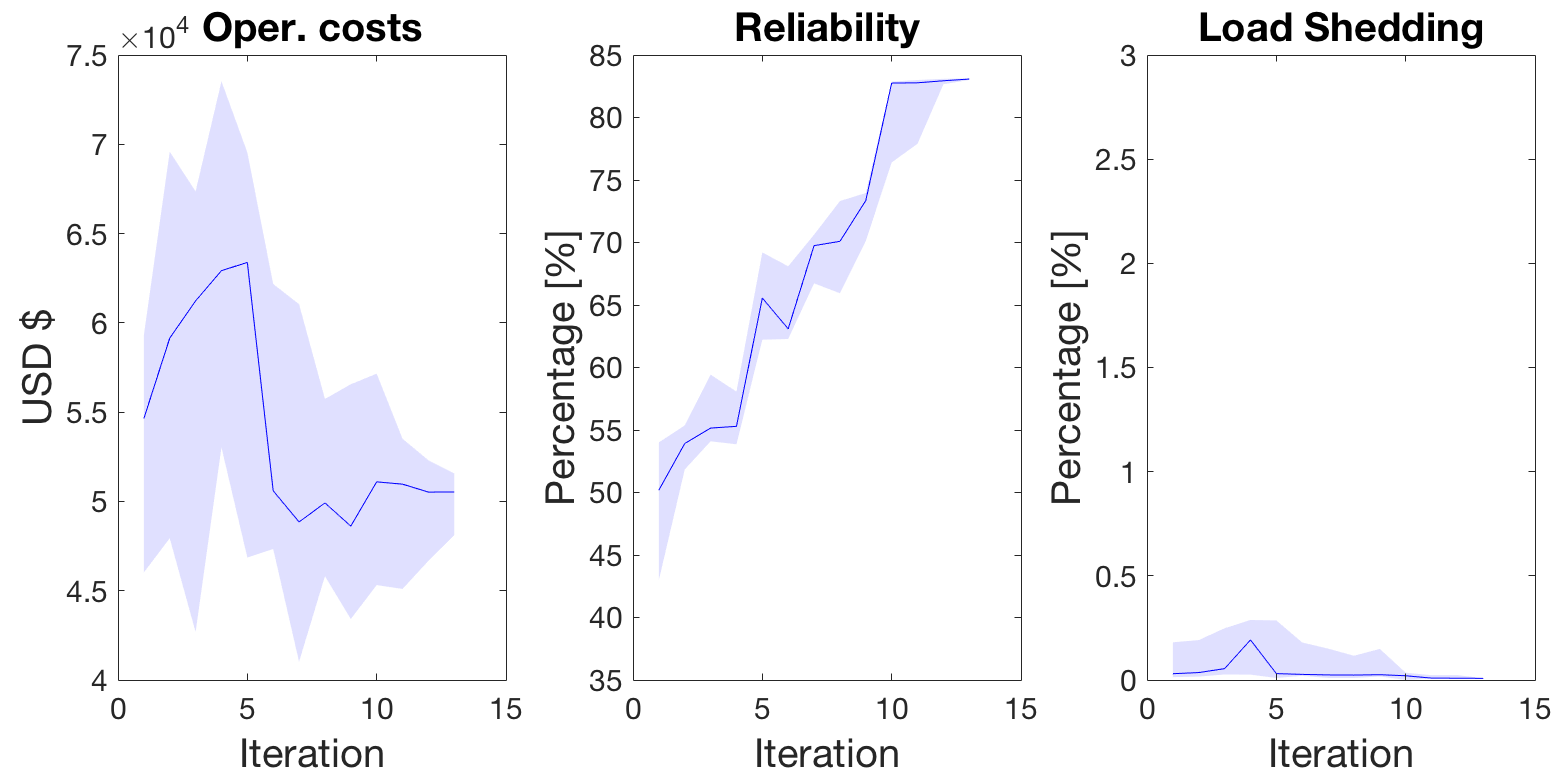}
		\caption{IEEE-RTS96}
	\end{subfigure}
	
	\caption[Two numerical solutions]{Convergence of the Cross Entropy method. {Plotted are} medians with upper and lower quartiles of three metrics for the top CE percentile: operational costs (redispatch, wind curtailment and unit re-commitment), average reliability, and average load shedding .}
	\label{fig:convergence}
\end{figure}

{We also visualize the convergence in the space of outage schedules in Fig.~\ref{fig:planimage} via gray-level-mapped matrices.} The rows of these matrices denote the transmission lines and their columns denote the outage moments. For a given {entry}, the gray-level corresponds to the {relative intensity of outages selected for the specific line-month combination.} The initial CE iteration begins with uniformly-drawn candidate outages moments, followed by convergence towards a single solution. In the case of RTS96, the zonal time-allocation can be seen in the form of three shaded blocks of entries, with the three interconnection outages in the form of three shaded independent rows.

\begin{figure}
	\centering
	\begin{subfigure}[]{0.5\textwidth}
		\centering\includegraphics[width=0.99\linewidth]{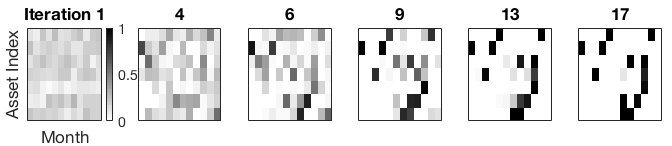}\vspace*{-2mm}
		\caption{IEEE-RTS79}\vspace*{1mm}
	\end{subfigure}
	
	\begin{subfigure}[b]{0.5\textwidth}
		\centering\includegraphics[width=0.99\linewidth]{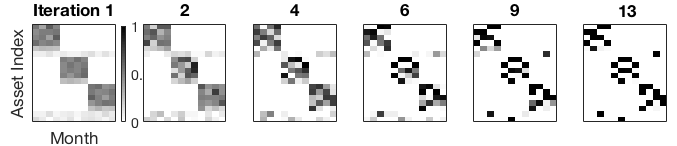}\vspace*{-2mm}
		\caption{IEEE-RTS96}
	\end{subfigure}
	\caption[Two numerical solutions]{{Gray-scale matrices representation of  drawn outages along several of the CE iterations, demonstrating convergence to a single, optimal schedule. The rows of each matrix denote the transmission lines and their columns denote the outage month.}}
	\label{fig:planimage}
\end{figure}

To demonstrate the efficacy of the optimal outage schedule, we compare it with arbitrary candidate outage schedules for RTS96. We do so by generating $100$ random schedules that comply with the zonal time allocation, and evaluating them as described {throughout this work}. We then perform $10$ additional evaluations of {our solution to \eqref{eq:fullMidTermFormulation}}. Fig.~\ref{fig:100 vs 10}(a) displays operational cost, reliability, and load shedding histograms of the $110$ evaluated schedules. {Our} optimization solution consistently exhibits the lowest operational costs, highest reliability, and lowest load shedding. One exception to its optimality is a single random schedule that achieves reliability greater than $90\%$. Therefore, we also include a {scatter plot in Fig.~\ref{fig:100 vs 10}(b)}  to capture the reliability vs. load shedding tradeoff. It reveals that the aforementioned highly-reliable random schedule suffers from high load-shedding values, as opposed to our optimization solution.

\begin{figure}
	\begin{subfigure}[]{0.5\textwidth}
		\centering
		\includegraphics[width=0.99\linewidth]{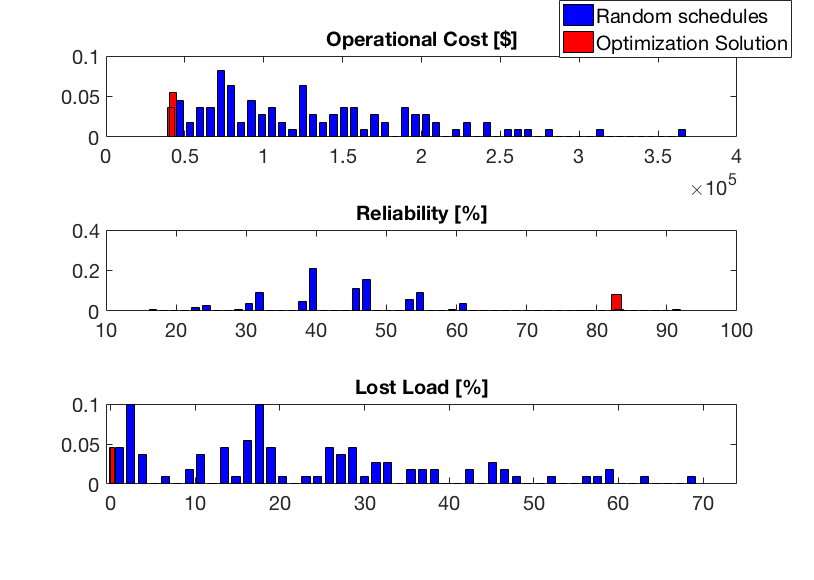}\vspace*{-7mm}
		\caption{Histograms of operational cost, reliability, and load-shedding.}
	\end{subfigure}
	\begin{subfigure}[b]{0.5\textwidth}
			\centering\vspace*{3mm}
		\includegraphics[scale=0.3]{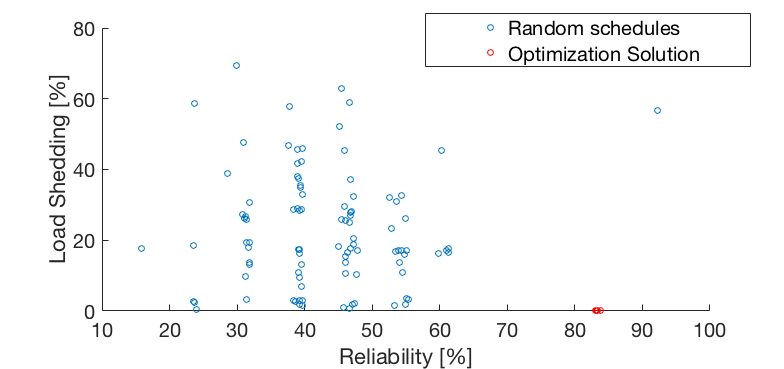}
		\caption{Reliability vs. load-shedding scatter plot for IEEE-RTS96}
	\end{subfigure}
	
	\caption[Two numerical solutions]{A comparison of $100$ arbitrary candidate outage schedules to $10$ instances of our optimization solution. In (a), the dominance of our solution is shown in all three inspected metrics. There is one single exception to its optimality: a random schedule with $>90\%$ reliability. However, in (b), where the reliability vs. load-shedding tradeoff is {depicted}, this single schedule is shown to suffer from high load shedding.}
	\label{fig:100 vs 10}
\end{figure}


\section{Conclusion} \label{sec:conclusion}

The power system infrastructure is ageing, and its maintenance is becoming more and more costly and complex. {This calls for new sophisticated outage scheduling tools, that will handle uncertainty and coordination with operations}.
The scenario assessment framework introduced in this work enables detailed evaluation of {implicit intricate} implications an outage schedule inflicts on a power system. We harness the power of machine learning and distributed computing {to tractably perform} multiple schedule assessments in parallel. We also wrap it in an optimization framework that finds convincingly high-quality schedules. {An additional, straightforward application of the methodologies introduced here is assessment of} a predefined maintenance schedule considered by experts.

{The focus of this work is in the probabilistic framework and hierarchical methodologies. Nevertheless, we believe it enables} gaining new insights for both academic networks and more complex real-world test-cases.

The proposed framework is flexible and can be adapted to different practical cost functions and reliability criteria.

\bibliographystyle{IEEEtran}
\bibliography{OS_transactions}
\vspace{-1cm}
\begin{IEEEbiographynophoto}{Gal Dalal}
	obtained his BSc in 2013 in electrical engineering from Technion, Israel. Since then he has been a PhD student in Technion. His research interests lie on the intersection between power systems and probabilistic optimization methods, machine and reinforcement learning.
\end{IEEEbiographynophoto}
\vspace{-1.3cm}
\begin{IEEEbiographynophoto}{Elad Gilboa}
 received a B.A in electrical and computer engineering and M.E. in biomedical engineering respectively in 2004 and 2009, from Technion, Israel, and a PhD in electrical engineering in 2014 from Washington University in St. Louis, USA. After which he did postdoctoral work in the Machine Learning Group at the Technion. His research includes machine learning and statistical single processing, with focus on making these methods scalable to real world problems.
\end{IEEEbiographynophoto}
\vspace{-1.3cm}
\begin{IEEEbiographynophoto}{Shie Mannor}
	received the B.Sc. degree in electrical engineering, the B.A. degree in mathematics, and the Ph.D. degree in electrical engineering from the Technion-Israel Institute of Technology, Haifa, Israel, in 1996, 1996, and 2002, respectively. From 2002 to 2004, he was a Fulbright scholar and a postdoctoral associate at M.I.T. He was with the Department of Electrical and Computer Engineering at McGill University from 2004 to 2010 where he was the Canada Research chair in Machine Learning. He has been with the Faculty of Electrical Engineering at the Technion since 2008 where he is currently a professor.
	His research interests include machine learning and pattern recognition, planning and control, multi-agent systems, and communications.
\end{IEEEbiographynophoto}
	\vspace{-1.2cm}
\begin{IEEEbiographynophoto}{Louis Wehenkel}
	graduated in Electrical Engineering (Electronics) in 1986 and received the Ph.D. degree in 1990, both from the University of Li\`ege, where he is full Professor of Electrical Engineering and Computer Science. His research interests lie in the fields of stochastic methods for systems control, optimization, machine learning and data mining, with applications in complex systems, in particular large scale power systems planning, operation and control, industrial process control, bioinformatics and computer vision.
\end{IEEEbiographynophoto}

\appendices

\section{Short-term and Real-time Mathematical Formulations} \label{sec:unit_commitment}
In the case of DC power flow, voltage magnitudes and reactive powers are eliminated from the problem, and real power flows are modeled as linear functions of the voltage angles \cite{zimmerman2011matpower}. This results in a mixed integer-linear program (MILP) that can be solved efficiently using commercial solvers \cite{CPLEX}. The full unit-commitment formulation is the following.

\begin{subequations}  \label{eq:unit_commitment}
	\begin{align}
	& u_p^* = \argmin_{u_p \in \mathcal{U}_p(u_m, y_s)}   C_{p}(u_m,u_p) = \argmin_{\alpha,\Theta,P_{g,t},WC,LS} \nonumber \\
	& \sum_{t=1}^{T{d.a}} \left[\sum_{i=1}^{n^g_d}\left(\alpha_{t}^if_P^i(P_{g,t}^i) + \alpha_{t}^i(1-\alpha_{t-1}^i)SU_i(t_{\text{off}}^i(\alpha,t) \right)\right.   \nonumber \\
	&\quad \quad \quad + \left. \sum_{iw=1}^{n^g_w}WC_{t}^{iw}\cdot C_{WC} + \sum_{ib=1}^{n^b}LS_{t}^{ib}\cdot VOLL \right], \label{eq:objective}\\
	&\text{subject\ to}  \\
	& g_{P,t}^l(\Theta^l,\alpha,P_g)=~B_{\text{bus}}^l \Theta_{t}^l + P_{BUS,\text{shift}}^l + {D}_{d.a,t} \label{eq:power_balance} \\
	& ~~+G_{sh} - LS_{t}  - ({W}_{d.a,t}-WC_{t})  - C_g (\alpha_{t}.* P_{g,t}) = 0, \nonumber\\
	& h_{f,t}^l(\Theta_{t}^l) =  B_f^l \Theta_{t}^l + P_{f,\text{shift}}^l - F_{max}^l \leq 0, \\
	& h_{t,t}^l(\Theta_{t}^l) = B_f^l \Theta_{t}^l - P_{f,\text{shift}}^l - F_{max}^l \leq 0, \label{eq:to_line_limits} \\
	& \theta_i^{\text{ref}} \leq \theta_{i,t}^l \leq \theta_i^{\text{ref}}  \quad i \in {\cal I}_{\text{ref}}, \label{eq:angle_limits}\\
	& \alpha_{t}^iP_g^{i,\text{min}} \leq P_{g,t}^i \leq \alpha_{t}^iP_g^{i,\text{max}} \quad  i=1,\dots,n^g_d,\\
	& 0 \leq WC_{t}^{iw} \leq {W}_{d.a,t}^{iw} \quad  iw=1,\dots,n^g_w,\\
	& 0 \leq LS_{t}^{ib} \leq {D}_{d.a,t}^{ib} \quad  ib=1,\dots,n^b, \label{eq:LS_limit}\\
	& t_{\text{off}}^i(\alpha,t) \geq  t_{\text{down}}^i \quad  i=1,\dots,n^g_d, \label{eq:min_down}\\
	& t_{\text{on}}^i(\alpha,t) \geq  t_{\text{up}}^i\quad   i=1,\dots,n^g_d, \label{eq:min_up}\\
	&l=0,1,\dots,n^l_t, \label{eq:n-1 constraint}\\
	&t=1,\dots,T_\text{d.a}.
	\end{align} 
\end{subequations}

Formulation \eqref{eq:unit_commitment} generally supports ensuring the N-1 security criterion via \eqref{eq:n-1 constraint}. Specifically, in our simulations we only considered the N-0 case as noted in the paper body, which is obtained by replacing \eqref{eq:n-1 constraint} with $l=0.$ The formulation's components are explained as follows.

\begin{itemize}
	\item $l$ denotes index of a transmission line that is offline. $l=0$ denotes all lines are connected and online. lines undergoing an outage are excluded from $n^l_t.$ 
	\item $\alpha \in \{0,1\}^{n_d \times T_\text{d.a}}$ denotes commitment (on/off) status of all dispatchable generators at all time-steps.
	\item $\Theta \in [-\pi,\pi]^{ n_b \times (n_l+1) \times T_\text{d.a}}$ denotes  voltage angle vectors for the N-1 network layouts at all time steps.
	\item $P_g \in \mathbb{R}_+^{n_d \times T_\text{d.a}}, WC \in \mathbb{R}_+^{n_w \times T_\text{d.a}}, LS \in \mathbb{R}_+^{n_b \times T_\text{d.a}}$ denote dispatchable generation, wind curtailment and load shedding decision vectors, with $f_P, C_{WC}, VOLL$ being their corresponding prices.
	\item $t_{\text{down}}^i, t_{\text{up}}^i$ denote minimal up and down time limits for generator $i$, after it had been off/on for $t_{\text{off}}^i$/$t_{\text{on}}^i;$ the latter are functions of $\alpha$ and $t$, as depicted in $\eqref{eq:objective}.$
	\item $SU_i(t_{\text{off}}^i(\alpha,t))$ denotes start-up cost of dispatchable generator $i$ after it had been off for $t_{\text{off}}^i$ time-steps.
	\item $g_{P,t}^l(\Theta^l,\alpha,P_g)$ denotes the overall power balance equation for line $l$ being offline.
	\item $B_{\text{bus}},P_{BUS,\text{shift}}$ denote nodal real power injection linear  coefficients.
	\item $B_f,P_{f,\text{shift}}$ denote linear coefficients of the branch flows at the \emph{from} ends of each branch (equal minus of the \emph{to} ends, due to the lossless assumption).
	\item $G_{sh}$ denotes a vector of real power consumed by shunt elements.
	\item $C_g$ denotes generator-to-bus connection matrix, where $(\alpha_{t}.* P_g)$ denotes the dot-product of the two vectors.
	\item $F_{max}$ denotes line flow limits.
	\item ${\cal I}_{\text{ref}}$ denotes the set of indices of reference buses, with $\theta_i^{\text{ref}}$ being the reference voltage angle.
	\item $P_g^{i,\text{min}},P_g^{i,\text{max}}$ denote minimal and maximal power outputs of generator $i$.
\end{itemize}

Furthermore,
\begin{itemize}
	\item \eqref{eq:power_balance}-\eqref{eq:to_line_limits} ensure load balance and network topology constraints;
	\item \eqref{eq:angle_limits}-\eqref{eq:LS_limit} restrict the decision variables to stay within boundaries. Namely, voltage angle limits, generator minimal and maximal power output range, wind curtailment and load shedding limits; and
	\item \eqref{eq:min_down}-\eqref{eq:min_up} bind the different time steps to follow generator minimal up and down time thermal limits.
\end{itemize}

As for the real-time optimization problem, it results in a formulation similar to the operational planning formulation in \eqref{eq:unit_commitment}. However, it involves the following adaptations:
\begin{itemize}
	\item it is individually solved per each time hour, instead of a whole day-ahead horizon $T_{\da}$;
	\item the on/off commitment schedule is no longer a decision variable, rather it is retrieved from $u_s^*$ and is set as a constraint in real-time operation;
	\item wind power and load forecasts $\hat{W}_{\da},\hat{D}_{\da}$ for the $T_{\da}$ time-steps are replaced with their actual realizations $W_t,D_t$; 
	\item a symmetric re-dispatch cost is added to the objective:  $\sum_{i=1}^{n^g_d}\alpha_{t'}^{*,i}|f_P^i(P_{g,t'}^{*,i})-f_P^i(P_{g,t'}^i)|$. It compares generation planned in the day-ahead plan $P_{g,t'}^{*,i}$ and the actual realized power consumed in real-time $P_{g,t'}^i$.
\end{itemize}
Having as an input the full realized state $y_\RT=\s{t}$ (either by witnessing it in real time, or by sampling future realizations), we  solve the real-time decision problem and obtain the voltage magnitude and angle at all network nodes. It can then be potentially used to evaluate related phenomenas, such as aggregated stress effect on equipment failure. However, we leave such considerations to future research.

%
%
%
%
%
%
\section{Assumed Stochastic Processes}  \label{sec:transition_model}
The Bayesian factorization approach, on which our scenario sampling methodology relies, follows the following decomposition of the probability of state $\s{t}.$
\[\prob{\s{}}=\prob{y_{\RT}|y_s,y_m}\prob{y_s|y_m}\prob{y_m},\]
where the time index was stripped away for brevity. 

Notice that since each of the real-time and short-term processes are conditioned on higher levels in the hierarchy, the state sequence $\s{t}$ is a stationary Markov process; i.e., 
\[ \prob{\scenario{}}=\prob{\s{0}} \cdot \prob{\s{1}|\s{0}}\ldots \prob{\s{T}|\s{T-1}},\] where $\prob{\s{t+1}|\s{t}}$ is a stationary state transition probability.

The stochastic processes in the system are the daily and hourly wind power produced in the wind generators $W_\da$ and $W_t$, respectively; the daily and hourly load process $D_\da$ and $D_t$, respectively; and the topology of the network $\text{top}_t$, used for the generation of training and test sets of the UCNN algorithm. We now provide details on the models used for these  stochastic components, along with the data and test-cases they are based on.

\subsubsection{Wind power distribution}
Wind generation capacities are taken from \cite{UW_website}, along with their daily mean profile. 
In addition, a monthly wind profile is adopted from \cite{klink1999trends}. 
The wind process mean $\mu_w(t)$ is obtained from the formula \[\mu_w(t) = \mu_w(t_D)\cdot p_{w,\text{annual}}(t_M),\] where $\mu_w(t_D) \in \mathbb{R}_+^{n_w}$ is the daily wind mean profile at time-of-day $t_D$, and $p_{w,\text{annual}}(t_M) \in [0,1]$ is the monthly wind profile relative to its peak at month $t_M$ of the year. 

Daily wind generation process $W_\da$ is multivariate normal: \[W_\da \sim \mathcal{N}\left(\mu_w(t),diag((p_{w,\sigma}\cdot \mu_w(t))^2)\right)\] where $p_{w,\sigma} \in [0,1]$ is a constant ($=0.15$) that multiplies the mean $\mu_w(t)$, to obtain a standard deviation that is a fixed fraction of the mean. $diag(x)$ is a square diagonal matrix, with the elements of $x$ as its diagonal, assuming wind generators to be uncorrelated.  $W_\da$ is truncated to stay in the range between $0$ and the generator's capacity.

Hourly wind generation  $W_t$ is assumed to be a biased random walk, with expectation $W_\da$; i.e.,  the real-time wind process is following the daily forecast up to some accumulated prediction error: 
\begin{align}
&W_t = W_\da(t) + \delta_t, \label{eq:wind_evolution}\\
&\delta_{t+1} = \delta_t + \epsilon_t, \label{eq:delta_evolution}
\end{align} 
where $\epsilon_t$ is Gaussian noise, distributed $\epsilon_t\sim {\mathcal N}(0,0.005\cdot W_\da(0)).$

\subsubsection{Load distribution}
Daily load $D_\da$ is assumed to follow a  distribution similar to the daily wind distribution $W_\da$, with the same formula containing peak loads and daily profiles for each bus $\mu_d(t_D) \in \mathbb{R}_+^{n_b}$ with values taken from \cite{UW_website}.
Fraction of mean for standard deviation is set to be $p_{d,\sigma}=0.02$.

Equivalently, hourly load process $D_t$ follows its day-ahead forecast $D_\da$ up to some accumulated error, as depicted in  \eqref{eq:wind_evolution} and \eqref{eq:delta_evolution}; there, the noise distribution is  $\epsilon_t\sim {\mathcal N}(0,0.001\cdot D_\da(0)).$ 

\subsubsection{Outage distribution} 
Generation of the training and test sets for UCNN proxy involves sampling UC inputs $\{x_i\}_{i=1}^n$ from distribution $\mathbb{P}_X(x)$. Our sampling technique is based on the following factorization of the random vector $x$: daily wind power $W_{\text{d.a}}$ and daily load $D_{\text{d.a}}$ are statistically independent conditioned on the month of the year (which is drawn uniformly first), whereas daily network topology $\text{top}_{\text{d.a}}$ is independent of them both. Each independent component is thus sampled from its marginal distribution.	
Section~\RNum{6} lists the choice of transmission lines where outages are considered. Sampling of daily topology $\text{top}_{\text{d.a}}$ is done  uniformly out of the combinatorial outage set.

\begin{figure*}
	\includegraphics[scale=0.42]{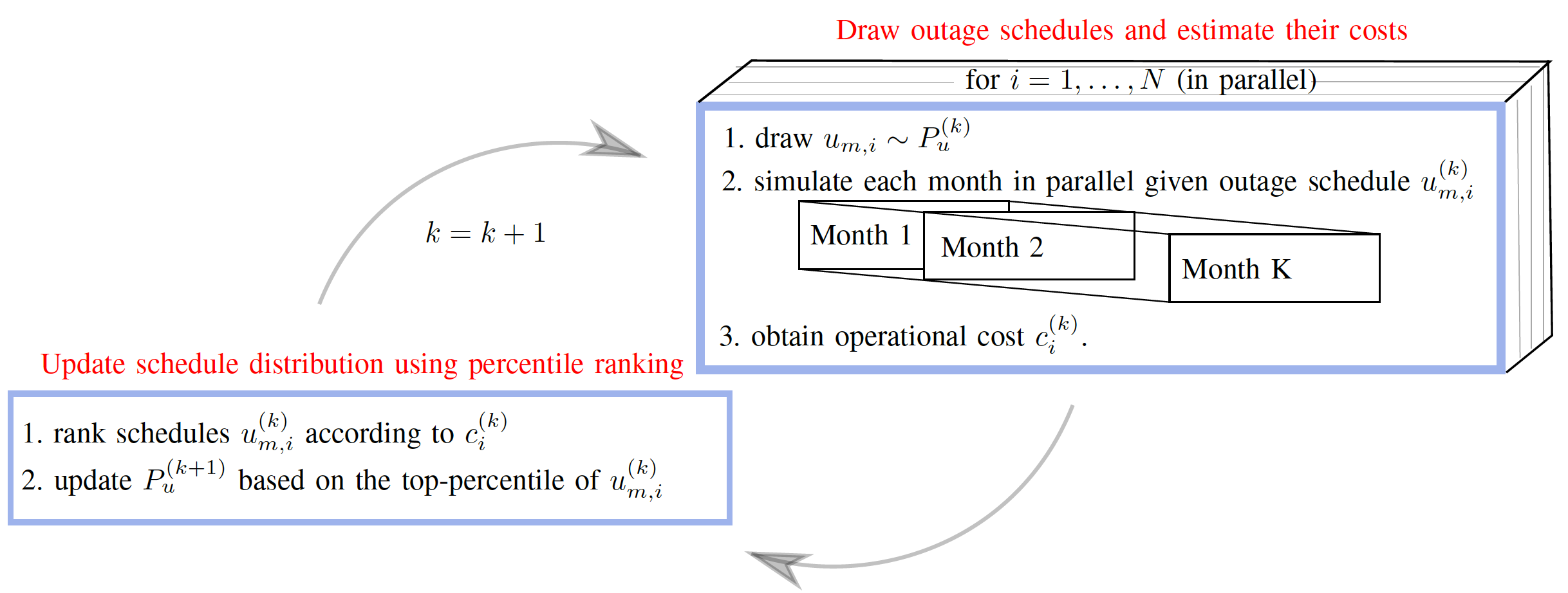}
	\caption{Cross Entropy: visualization.}
	\label{fig:CE}
\end{figure*}

\section{Distributed Cross Entropy Optimization}
We solve our outage scheduling formulation using the Cross Entropy (CE) optimization meta-heuristic \cite{de2005tutorial}. It is a randomized percentile optimization method that has shown to be useful for solving power system combinatorial problems \cite{ernst2007cross}. In CE, a parametric distribution $P_u$ is maintained over the solution space ${\cal{U}}_m$. Per each iteration $k$ until convergence of $P_u^{(k)}$, it performs consecutive steps of
\begin{enumerate}
	\item drawing $N$ candidate outage schedules $u_{m,i}^{(k)} \sim P_u^{(k)},~i=1,\dots,N,$ and evaluating their respective costs in parallel, based on simulated sampled scenario set $\hat{{\cal  Z}}^{(k)}$ (constructed with our Bayesian hierarchical window scenario sampler and decomposed at the monthly level):
	$$c_i^{(k)} = \frac{1}{|\hat{{\cal  Z}}^{(k)}|}\sum_{\hat{Z} \in \hat{{\cal  Z}}^{(k)}}\sum_{{\s{t} \in \hat{Z}}} {C}(\s{t},u_{m,i}^{(k)}, u_s^*, u_\RT^*);\mbox{ and}$$ 
	
	\item updating the parametric solution distribution $P_u^{(k)}$ based on the cheapest $0.15$-percentile of $c_i^{(k)}.$
\end{enumerate} 
This iterative process is visualized in Fig.~\ref{fig:CE}.

Specifically, in our experiments the outage scheduling horizon is one year with monthly candidate outage moments. Therefore, $\actionm$ is a binary matrix; i.e., $\actionm \in \actionSetm = \{0,1\}^{|{\cal L}| \times 12},$  where ${\cal L}$ is the set of transmission lines for which outages ought to be performed. Entry $\actionm(l,m)=1$ indicates a scheduled outage of line $l$ during month $m$. 

We thus represent the CE distribution $P_u^{(k)}$ with a matrix of size $|{\cal L}| \times 12$ whose entries $p^{(k)}_{l,m} \in [0,1]$ depict outage likelihood. At iteration $k=0,$ these are all initialized to $0.5$. As explained in the experiments section, according to the outage lists for IEEE-RTS79 and IEEE-RTS96 we need to schedule either $1$ or $2$ outages per each line in ${\cal L}$, depending on the line. Thus, per each row $p^{(k)}_{l,:} \in [0,1]^{12}$ (corresponding to line $l$) we respectively draw $1$ or $2$ entries out of the $12$ candidate entries. This per-row sampling is performed by 
drawing one out of the $\binom{12}{1}$ or alternatively $\binom{12}{2}$ entry combinations based on their proportional probability, calculated using matrix $[p^{(k)}]_{l,m}$. The first step of the CE algorithm is thus to iterate the above procedure $N$ times for sampling $u_{m,i}^{(k)},~i=1,\dots,N$.

The second step of the CE algorithm is to update $P_u^{(k)}$ as follows. Let ${\cal C} \subset \{1,\dots,N\}$ be the set of indices of the cheapest $0.15$-percentile of costs $c_i^{(k)}.$ Then, we update the matrix $[p^{(k+1)}]_{l,m} = \frac{1}{|{\cal C}|} \sum_{j \in {\cal C}} u^{(k)}_{m,j};$ i.e., the entries of the matrix representing $P_u^{(k+1)}$ are set to be the average of the top solutions' entries. 

Lastly, our criterion for convergence is when the entropy of  $[p^{(k+1)}]_{l,m}$ drops below some small $\epsilon>0.$ This occurs when all entries are sufficiently close to either $0$ or $1$.

\end{document}